\documentclass[12pt, preprint]{aastex}

\usepackage{amsmath}
\usepackage{amssymb}

\newcommand{\be}{\begin{equation}}
\newcommand{\ee}{\end{equation}}
\newcommand{\bea}{\begin{eqnarray}}
\newcommand{\eea}{\end{eqnarray}}

\newcommand{\fzv}{f_0\left (z,\,v_z\right )}
\newcommand{\fzvp}{f_1\left (z,\,v_z,\,t\right )}

\begin{document}

\title{Galactoseismology:  Discovery of Vertical Waves in the Galactic Disk}

\author{
Lawrence M. Widrow$^{1}$,
Susan Gardner$^{2}$,
Brian Yanny$^{3}$,
Scott Dodelson$^{3,4,5}$,
Hsin-Yu Chen$^4$
}
\affil{
${}^{1}$Department of Physics, Engineering Physics, and Astronomy, Queen's University, Kingston, ON, K7L 3N6, Canada\\
${}^{2}$Department of Physics and Astronomy, University of Kentucky, Lexington, KY 40506-0055\\
${}^{3}$Fermi National Accelerator Laboratory,  Batavia, IL 60510\\
${}^{4}$Department of Astronomy and Astrophysics, The University of Chicago, Chicago, IL 60637\\
${}^{5}$Kavli Institute for Cosmological Physics, Chicago, IL 60637
}

\begin{abstract}
  We present evidence for a Galactic North-South asymmetry in the
  number density and bulk velocity of solar neighborhood stars.  The
  number density profile, which is derived from main-sequence stars in
  the Sloan Digital Sky Survey, shows a (North$-$South)/(North$+$South)
  deficit at $|z|\simeq 400\,{\rm pc}$ and an excess at $|z|\sim
  800\,{\rm pc}$.  The bulk velocity profile, which is derived from
  the Sloan Extension for Galactic Understanding and Exploration,
  shows a gradual trend across the Galactic midplane as well as
  smaller-scale features.  We speculate that the North-South
  asymmetry, which has the appearance of a wavelike perturbation, is
  intrinsic to the disk.  We explore the physics of this phenomenon
  through an analysis of the linearized Boltzmann and Poisson
  equations and through one-dimensional simulations.  The perturbation
  may be excited by the passage of a satellite galaxy or dark matter
  subhalo through the Galactic disk, in which case we are witnessing a
  recent disk-heating event.

\end{abstract}



Disk galaxies are dynamic systems that can develop bars, spiral
structure, and warps.  They tidally disrupt satellite galaxies and dark
matter subhalos, a process that leaves behind streams of stars and
dark matter.  Likewise, satellites continually perturb the disk of their
host galaxy.

Despite the existence of these time-dependent phenomena, the
assumption that galaxies are in equilibrium (i.e., stationary in the
potential) has been used extensively to interpret certain types of
observations.  Best known, perhaps, are attempts to infer the surface
density and vertical force near the Sun from the kinematics of stars in
the solar neighborhood.  \citet{oort32} pioneered this program and
elements of his original method are present in virtually all studies
of this type.  Approaches to the so-called Oort problem generally
involve an estimate of the distribution function for solar
neighborhood stars, $f\left (z,\,v_z\right )$, where $z$ is the
position relative to the midplane of the Galaxy and $v_z$ is the
vertical velocity.  For stars in equilibrium, and not too far above
the midplane, this distribution is a function of the effective
vertical energy per unit mass, $E_z = v_z^2/2 + \Psi(z)$, where
$\Psi(z)$ is the gravitational potential in the solar neighborhood.
Numerous researchers including \citet{Bahcall84}, \citet{Kuijken89b,
  Kuijken89a}, \citet{flynn94}, and \citet{Holmberg00} have extended
and improved upon Oort's method.

In this Letter, we estimate the number density of stars with both
photometric and astrometric observations from SDSS-DR8, the Eighth
Data Release of the Sloan Digital Sky Survey \citep{aihara11}.  We
also use a smaller spectroscopic data set from SEGUE, the Sloan
Extension for Galactic Understanding and Exploration \citep{yanny09},
to estimate the bulk velocity and velocity dispersion.  While our
motivating interest was in revisiting the Oort problem, our aim in
this letter is to demonstrate the existence of a North-South asymmetry
in both the spatial density and velocity distribution of solar
neighborhood stars.  We explore the hypothesis that the asymmetry
reflects a coherent wavelike perturbation, which is intrinsic to the
disk.  A theoretical discussion, together with a simple
one-dimensional simulation, reveals that perturbations of this type
are entirely natural.

We begin with a discussion of the photometric data set.  Since our
primary interest is in differences in the stellar distribution North
and South of the Galactic midplane, we choose matched regions from the
survey above and below the Sun.  Specifically, we select stars with
Galactic coordinates $100^\circ < l < 160^\circ$ and $54^\circ < |b| <
68^\circ$ that reside within a perpendicular distance of $1\,{\rm kpc}$
from the line connecting the North and South Galactic poles.  We
determine dereddened apparent magnitudes in $g$, $r$, and $i$ using
the extinction maps of \citet{schlegel98}.  In doing so, we assume
that all of the stars are beyond the dust, which is thought to lie at
$|z|<125\,{\rm pc}$ \citep{marshall06}.  Only stars with $14 < r < 21$
are included in order to avoid problems with saturation and
photometric errors.  We calculate absolute $r$-band magnitudes, $M_r$,
using $r-i$ colors and the photometric parallax relation from
\citet{juric08} as well as $g-i$ colors and the corresponding relation
from \citet{ivezic08}.  The selection function, which depends on the
geometry of the survey and our brightness cuts, is used to convert the
observed number of stars to stellar number density with height.  A
detailed discussion of the selection function and various systematic
effects will be discussed in a forthcoming publication.

In Figure 1, we plot the number density $n(z_{\rm obs})$, where
$z_{\rm obs}\equiv z - z_\odot$ and $z_\odot$ is the vertical position
of the Sun.  We consider stars with $0.6 < r-i < 1.1$ and compute
distances using the photometric parallax relation from
\citet{juric08}.  In this $r-i$ range, we observe all stars
(approximately 300K) with $0.2\,{\rm kpc} \la |z_{\rm obs}| \la
1.6\,{\rm kpc}$.  The density profile shows the well-documented excess
over a single exponential for $|z|\ga 1\,{\rm kpc}$ \citep{gilmore83,
  reid93, juric08}.  We fit a two-component model
\begin{equation}\label{eq:rhomodel}
n(z_{\rm obs}) ~=~ n_0\left ({\rm sech}^2\left (\frac{z_{\rm obs}+z_\odot}{2H_1}\right ) + f
{\rm sech}^2\left (\frac{z_{\rm obs}+z_\odot}{2H_2}\right )\right )
\end{equation}
using $\chi^2$-minimization \citep{press92} and find $n_0 = \left
(4.22\pm 0.04\right )\times 10^6\,{\rm kpc}^{-3}$, $H_1= \left (221\pm
3\right )\,{\rm pc}$, $H_2 = \left (582 \pm 12\right ){\rm pc}$, $f=
0.17\pm 0.01$, and $z_\odot = \left (15.4\pm 0.8\right )\,{\rm pc}$.
The residuals $\Delta$
$\equiv \left ({\rm data}-{\rm model}\right )/{\rm model}$
are largely an odd
function in $z$.  In the North, the data lie below the model for
$z\simeq 400\,{\rm pc}$ and above the model for $z\sim 800\,{\rm pc}$
while the situation is reversed in the South.  We note that while
equation 1 captures the general features of the measured $n(z_{\rm
  obs})$, the fit is relatively poor with $\chi^2\simeq 444$ for $47$
degrees of freedom.  We strongly rule out a symmetric disk model, even
when we assume some correlation of errors in adjacent color bins that
might arise from the conversion of colors to distances via the
photometric parallax relation.

A consequence of the $z$-odd residuals is that the model parameters are
sensitive to the range in $z$ considered.  In particular,
if we fit only data within $500\,{\rm pc}$, we find $z_\odot\simeq
39\,{\rm pc}$.  Thus, our two determinations of $z_\odot$ bracket the
typical values found in the literature \citep{juric08}.

To explore the density structure further, we consider the North-South
asymmetry, defined as $A\left (z>0\right )\equiv \left (n(z) -
n(-z)\right )/\left (n(z) + n(-z)\right )$.  In Figure 2a, we show $A$
for a series of color bins from $r-i=0.4$ to $r-i=1.4$.  In
constructing this figure, we use $z_\odot=15\,{\rm pc}$.  Note that
the probed range in $z$ depends on color; regions close to the
midplane are probed by the redder bins and vice versa.  Evidently, $A$
shows little color dependence.  In Figure 2b, we show $A$ for the
$r-i$ window used in Figure 1.  Also shown is the asymmetry obtained
if distances for these stars are computed using the $g-i$ photometric
parallax relation.  In Figure 2c, we compute the composite asymmetry
over all color bands in Figure 2a as well as the composite asymmetry
found for stars with $1.6 < g-i < 2.6$.  We also show the asymmetry
when we assume $z_\odot = 39\,{\rm pc}$.

The statistical significance of the North-South asymmetry may be
tested using the Kolmogorov-Smirnoff (K-S) test where
the statistic $D$ is defined as the maximum difference between
fractional cumulative distributions of stars North and South of the
Galactic midplane.  We carried out a K-S test for subsamples with different
numbers of stars $N$ and found that $D\simeq 0.035-0.055$ for
$N>9000$.  The probability that the stars North and South of the
midplane are drawn from the same distribution is $P\simeq \exp\left
(-D^2N\right )$ \citep{press92}.  Thus, with only a few percent of the
full data set we rule out the null hypothesis (North-South
symmetry distribution) at the $10^{-5}$ level.  For the full data set,
$P$ is vanishingly small.

In Figure 3 we show the two-dimensional asymmetry, $A_{2D}\left
  (R,\,z\right )$, where $R$ is the Galactocentric radius.  This
figure may be compared with Figure 26 from \citet{juric08}, who also
studied SDSS stars.  The asymmetry described in this letter appears to
be evident in the bottom right panel of their figure.

We now consider the spectroscopic sample, which comprises some 11,000
SEGUE stars.  The $\left (l,b\right)$-footprint for this sample is
slightly larger than the footprint for the photometric sample, though
the $\left (l,b\right )$ coverage is not uniform \citep{yanny09}.  We
include stars with $1.0 < g-i < 1.9$ that satisfy the stellar surface
gravity cut from \cite{lee08}, which restricts our sample to dwarfs.
Proper motions and radial velocities are converted to a $\left
(v_x,\,v_y,\,v_z\right )$ coordinate system using the standard
transformation laws \citep{johnson87}.  Errors in $v_z$ are typically
$6-11\,{\rm km\,s^{-1}}$ and result from uncertainties in proper
motions, radial velocities and distances.  We consider stars with
$|v_z| < 125\,{\rm km\,s^{-1}}$ so as to avoid contamination from
high-velocity outliers.  In Figure 4a we see a trend in $\langle
v_z\rangle$ across the midplane as well as small scale undulations.
These results suggest that there is a coherent motion of solar
neighborhood stars away from the Galactic midplane.  We note that
recently \cite{bochanski11} found evidence for variations in $\langle
v_z\rangle$ with M-star color, which is correlated with $z$.  Figure
4c shows an increase in $\sigma_z$ with $|z|$, which
may be attributed to an increase in the fraction of kinematically
``hotter'' thick disk stars as one moves away from the midplane.

Various systematic effects can produce trends in $\langle v_z\rangle$.
\citet{bond10} analysed SDSS-DR7 data and reported a difference of
$\sim 8\,{\rm km\,s^{-1}}$ between the bulk motions North and South of
the midplane; they attributed this difference to a systematic radial
velocity error for M dwarfs possibly due to the limited set of
velocity template spectra used in the SDSS cross correlation.  To
explore this issue, we consider red and blue subsamples.  The results,
shown in Figure 4b, do indeed indicate a stronger trend in the red
subsample, though a trend and small scale features are still evident
in the blue subsample.  Another example is a systematic offset in
distances, which would enter $\langle v_z\rangle$ through the proper
motions, especially at mid-Galactic latitudes \citep{schoenrich12},
though a large offset ($\sim 30\%$) is required to explain the trend
seen in Figure 4a {\it in toto}.  Clearly, a definitive statement on
the $z$-dependence of $\langle v_z\rangle $ will require improvements
in radial velocity and distance measurements along with better
coverage toward the Galactic poles.

A primary goal of the SDSS and SEGUE projects is to discover new
structure in the Galaxy's stellar halo.  \citet{juric08}, for example,
detected several new overdensities in the thick disk and stellar halo.
These features are usually interpreted as stellar debris from a
tidally disrupted satellite.  While it is tempting to seek a similar
explanation for the features described in this letter we explore a
different interpretation: {\it the structures seen in both
  the photometric and spectroscopic data represent coherent wavelike
  perturbations in the Galactic disk}.

Evidence for this interpretation can be found in the preceding
figures.  The model residuals (Figure 1) and North-South asymmetry
(Figure 2) extend above and below the Galactic midplane and exhibit a
wavelike structure.  Furthermore, the structures appear to run
parallel to the Galactic midplane, as seen in Figure 3.  In addition,
the bulk velocity exhibits a trend running through the
midplane and smaller-scale fluctuations.  Finally, and perhaps
most importantly, the asymmetry appears to be largely independent of color
(Figure 2a).  A strong color dependence would have suggested that the
asymmetry was due to different stellar populations North and South of
the midplane.  Indeed, \citet{schlafly10} note that F turn-off stars
in the South appear to be systematically redder in $g-r$ by $0.02$
mags than in the North.  However, these stars are typically located
$5-10\,{\rm kpc}$ from the Galactic plane, and therefore well beyond
the extent of the thin or thick disk stars studied here.

We now demonstrate that these features naturally arise in a perturbed
self-gravitating stellar disk. For simplicity, we ignore interstellar
matter and model the Galactic disk in the solar neighborhood as a
collisionless system of stars.  We further assume that the disk is
self-gravitating, though it is easy to include additional components
to the gravitational field.  Finally, since we work close to the
Galactic plane, we assume that we can decouple vertical motions of
disk stars from radial and azimuthal motions.

We consider small perturbations to the equilibrium solution
for a single-component, self-gravitating, isothermal plane.
The distribution function, density, and potential for the equilibrium
solution are
\begin{equation}
\label{eq:spitzerf}
\fzv ~=~\frac{\rho_0}{\left (2\pi\sigma^2\right )^{1/2}}
\exp{\left (-E_z/\sigma^2\right )}~,
\end{equation} 
\begin{equation}\label{eq:spitzerp}
\Psi_0(z)~=~2\sigma^2\ln\left (\cosh\left (\frac{z}{2z_d}\right )\right )~,
~~~~~~~
\rho_0(z)~=~ \rho_0{\rm sech}^2 \left
(\frac{z}{2z_d}\right )~,
\end{equation}
where $\rho_0 = \sigma^2/8\pi G z_d^2$
(\citep{spitzer42}, see, also \cite{binney08}, problem 4.21).
The linearized Boltzmann and Poisson equations are
\begin{equation}\label{eq:boltzmann}
\frac{\partial f_1}{\partial t}  ~+~ 
v_z\frac{\partial f_1}{\partial z} ~-~
\frac{\partial \Psi_1}{\partial z}
\frac{\partial f_0}{\partial v_z}~-~
\frac{\partial \Psi_0}{\partial z}
\frac{\partial f_1}{\partial v_z}
~=~ 0
\end{equation}
and
\begin{equation}
\label{eq:poisson}
\frac{\partial^2\Psi_1}{\partial z^2} ~=~ 4\pi G \rho_1~,
\end{equation}
where $\rho_1\left (z,\,t\right ) = \int dv_z \fzvp$.  Equation
\ref{eq:poisson} implies that the vertical force perturbation,
$F_1\equiv -\partial \Psi_1/\partial z$, is proportional to the
perturbed surface density, as defined from the Galactic midplane,
i.e., $F_1(z) = 4\pi G\left (\Sigma_1(z)-\Sigma_1(0)\right )$ where
$\Sigma_1\left (z\right )\equiv \int_{-\infty}^z dz\,\rho_1(z)$.
Moreover, we can integrate equation \ref{eq:boltzmann} over $v_z$ to
obtain the linearized continuity equation
\begin{equation}
\frac{\partial \rho_1}{\partial t} = -\frac{\partial}{\partial z}
\left (\rho\overline v\right )_1~,
\end{equation}
where $\left (\rho\overline v\right )_1\equiv \int dv_z\, f_1 v_z$.
The mean velocity
is therefore related to the surface density and vertical force:
\begin{equation}
\left (\rho\overline v\right )_1 ~\sim~
\frac{\partial\Sigma_1}{\partial t}
~\sim~
4\pi G \frac{\partial F_1}{\partial t}~,
\end{equation}
which implies a connection between the vertical perturbations
considered here and the Oort problem.

Equations \ref{eq:boltzmann} and \ref{eq:poisson} do not admit simple
plane wave solutions since the unperturbed model depends on $z$.
Nevertheless, the standard Jeans stability analysis suggests that
waves are described by the dispersion relation
\begin{equation} \label{eq:jeanslength}
k^2 \sim \frac{\omega^2}{\sigma^2} + \frac{4\pi G\rho}{\sigma^2}~,
\end{equation} 
where $k$ and $\omega$ are the wavenumber and angular frequency,
respectively.  A lower bound on the Jeans length, the minimum length
scale for gravitationally unstable modes, is obtained by setting
$\omega=0$ and $\rho$ and $\sigma$ equal to values characteristic of
the Galactic midplane.  We find $\lambda_J = 2\pi/k_J = \left
  (\pi\sigma^2/G\rho\right )^{1/2}\simeq 2\,{\rm kpc}$.  Since
$\lambda_J$ is much larger than the characteristic scale height of the
disk, we conclude that vertical perturbations are stable and behave
like pressure-supported waves.  Equation \ref{eq:jeanslength},
together with the fact that $\rho/\sigma^2$ decreases with $|z|$
implies that the wavelength for these perturbations will be somewhat 
less than $\lambda_J$ and will increase as one moves away from
the midplane.  These conclusions are consistent with the asymmetry
shown in Figure 2.

Since the equilibrium distribution function is even in both $z$ and
$v_z$, the characteristic modes of the perturbed system have definite
parity in both phase space coordinates. Any perturbation can be
modeled by displacing stars in position and velocity, that is, by
replacing the unperturbed phase space positions, $z_0$ and $v_0$ with
$z = z_0 + z_1(z_0)$ and $v_z = v_0 + v_1(z_0)$.  For the simplest
perturbation, all stars are displaced in the same direction.  We refer
to this type of perturbation as an $m=0$ mode since there are no nodes
in the displacement functions $z_1$ and $v_1$.  The $m=0$ mode amounts
to a local displacement of the stellar disk as a whole and affects any
determination of $z_\odot$ and $v_{z,\odot}$ from solar neighborhood
data.  (For observational evidence of $m=0$ modes or midplane
displacement in the molecular gas distribution of the Galaxy, see
\citet{nakanishi06}.)

The next simplest perturbation is an $m=1$ (breathing) mode where
displacements and peculiar velocities have opposite signs above and
below the midplane.  However, the density perturbation is an {\it
  even} function in $z$ and therefore $n(z)$ alone cannot reveal the
presence of an $m=1$ perturbation.  However, the $m=1$ mode would show
up as a North-South trend in the bulk velocity, which is consistent
with what we see in Figure 4.

The North-South asymmetry seen in Figure 2 is then likely a combination of
$m=2,\,4,\dots$ modes.  The local extrema in $A$ at $400$ and
$800\,{\rm pc}$ (and possibly $1200\,{\rm pc}$) imply nodes in $z_1$ at
the corresponding positions in $z$.  These higher order modes may also
be present in the velocity data.

We have numerically solved for the evolution of a one-dimensional
collisionless N-body system where each ``particle'' corresponds to an
infinite plane, which moves under the influence of the gravitational
field generated by the other planes.  The gravitational force on a
given plane is proportional to the difference between the number of
planes to the North and the number to the South.  (Similar simulations
were performed by \cite{sanchez11} in order to study vertical equilibrium
solutions of galactic disks.)


We chose an equilibrium model given by equations \ref{eq:spitzerf} and
\ref{eq:spitzerp} with $\sigma = 31\,{\rm km\,s^{-1}}$ and $z_d =
500\,{\rm pc}$.  These parameters imply a surface density at
$1.1\,{\rm kpc}$ of $71\,M_\odot\,{\rm pc}^{-2}$, which is the value
obtained by \citet{Kuijken89b}.  The initial perturbation is chosen to
roughly match the results found in Figures 1 and 4.  The evolution of
the system is shown in Figure 5. \footnote{See
  www.astro.queensu.ca/$\sim$widrow/OortExplorers.html for an animation
  of the simulation.} The simulations illustrate the interplay between
density and velocity perturbations.  Waves appear to reflect off the
low-density regions at high $z$, a phenomenon reminiscent of waves
propagating in the Sun, and decay over a period of $200-300\,{\rm
  Myr}$.

To conclude, we speculate on the origin of vertical perturbations in
galactic disks.  An isolated, self-gravitating disk can develop
non-axisymmetric structure, such as spiral arms and bars, through
gravitational instabilities (see \citet{binney08} and
\cite{sellwood10}).  On the other hand, the vertical modes are stable
and must be excited by some agent.  Of course, stellar disks of even
relatively isolated galaxies, such as the Milky Way, interact with
satellite galaxies and, presumably, halo substructure.  There is an
extensive body of literature on disk heating, dating back to the work
of \citet{lacey85}, who considered disk heating by supermassive black
holes, and \citet{velazquez99}, who considered disk heating by
satellite galaxies.  In addition, \citet{gauthier06} and
\citet{dubinski08} showed that satellites and halo substructure can
provoke spiral-type structures in a stellar disk and trigger bar
formation.  More recently, \citet{purcell11} proposed that the Sagittarius
dwarf galaxy was responsible for setting up the particular pattern of
spiral structure seen in the Milky Way.  Our conjecture is that the
vertical waves are another manifestation of disk heating and
disk-substructure interactions.

The observational, theoretical, and numerical analyses presented in
this paper can be extended to incorporate variations in the disk plane
and provide a more complete picture of the perturbed disk.  On the
observational front, future experiments, such as GAIA
\citep{perryman01, wilkinson05} and LSST \citep{ivezic08b}, which aim
to measure distances and velocities for some one billion stars, will
provide a picture of the phase space distribution of the Galaxy in
unprecedented detail.  These studies may well reveal a rich structure
of waves in the disk, and thereby open a new window into the dynamics
of the Galaxy.

\acknowledgements{We thank Evalyn Gates, Kristine Spekkens, John
  Dubinski, Ralph Sch\"{o}rich, and John Bochanski for useful
  conversations and the Aspen Center of Physics for its hospitality.
  We acknowledge use of SDSS-III data (http://www.sdss3.org).  LMW is
  supported by the Natural Sciences and Engineering Research Council
  of Canada.  SD is supported by the U.S.  Department of Energy,
  including grant DE-FG02-95ER40896, and by the National Science
  Foundation under Grant AST-090872.  SG acknowledges partial support
  from the U.S. Department of Energy under contract
  DE-FG02-96ER40989.}

\bibliographystyle{apj}

\begin{figure}
\epsscale{1.0}
\plotone{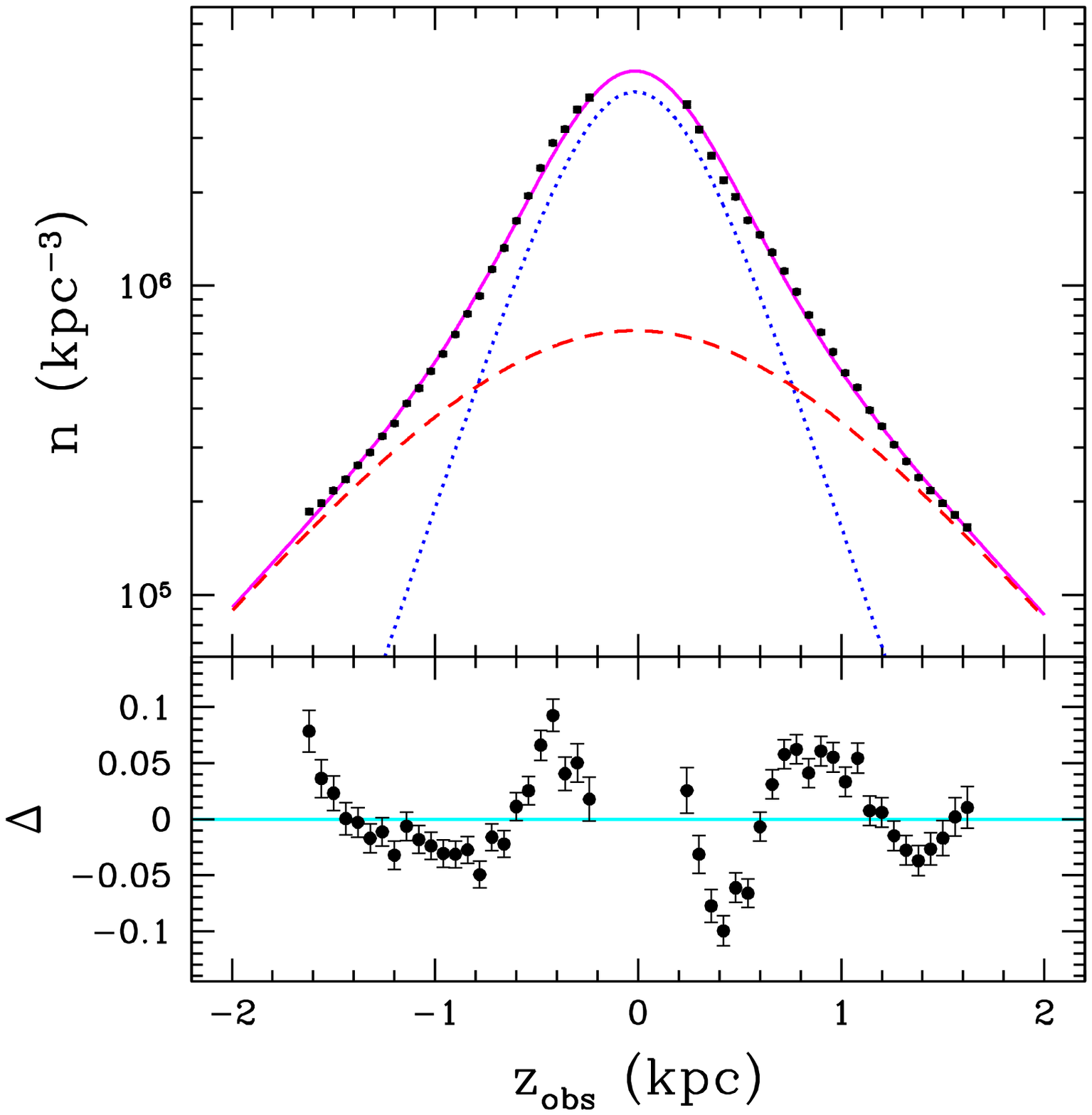}\label{fig:rhoz}
\caption{Number density $n$ as a function of distance from the Sun
$z_{\rm obs}$.  Black points are the data.  The magenta curve is our
model fit.  The dotted blue curve is the contribution from the thin
disk; the dashed magenta curve is the contribution from the thick
disk.  Lower panel shows residuals: $\Delta\equiv$ (data$-$model)/model.}
\end{figure}

\begin{figure}
\epsscale{1.0}
\plotone{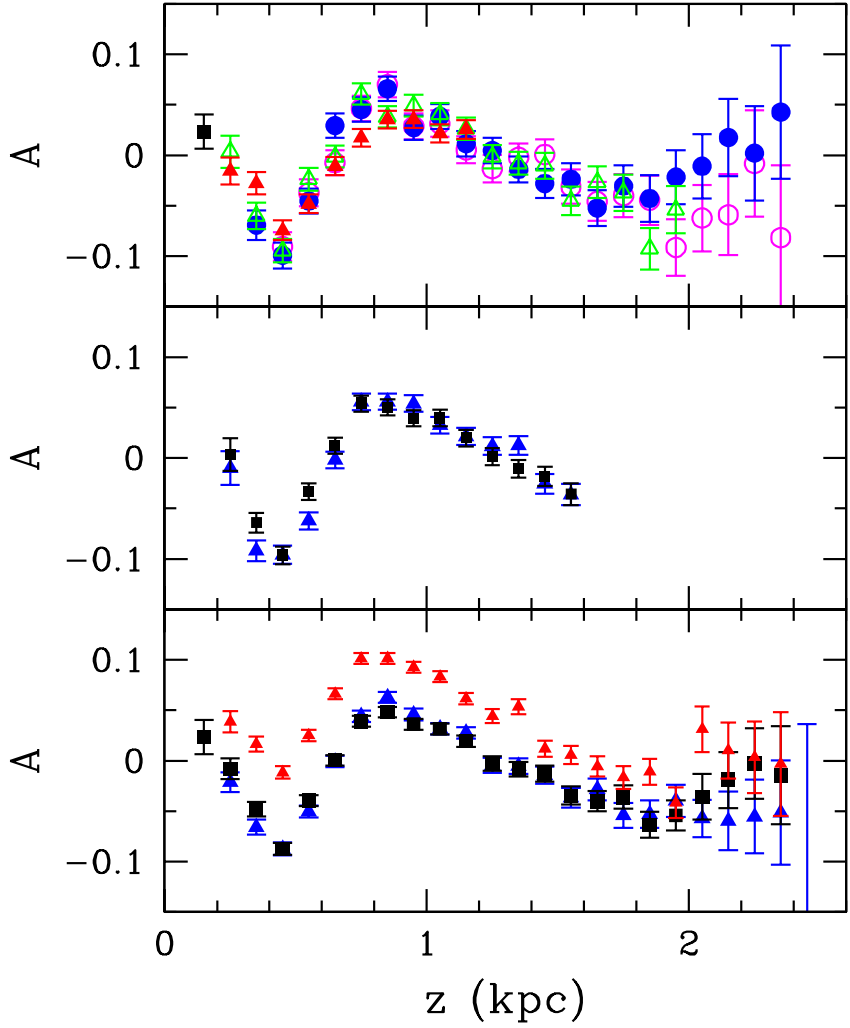}\label{fig:figure2a}
\caption{North-South asymmetry parameter $A$ as a function of $z$.
  Top panel shows $A(z)$ for $0.2$ mag color bins: black square --
  $r-i = \{1.2-1.4\}$; red filled triangle -- $r-i = \{1-1.2\}$; green
  open triangle -- $r-i = \{0.8-1\}$; blue filled circle -- $r-i =
  \{0.6-0.8\}$; magenta open circle -- $r-i = \{0.4-0.6\}$.  Middle
  panel gives the average $A$ for the color bins and $z$-range used in
  Figure 1.  Black points were generated using the photometric
  parallax relation from \cite{ivezic08}.  Bottom panel gives average
  $A$ over the full range in $z$ and $r-i$.  Red triangles assume
  $z_\odot=39\,{\rm pc}$ instead of $z_\odot=15\,{\rm pc}$.}
\end{figure}

\begin{figure}
\epsscale{1.0}
\plotone{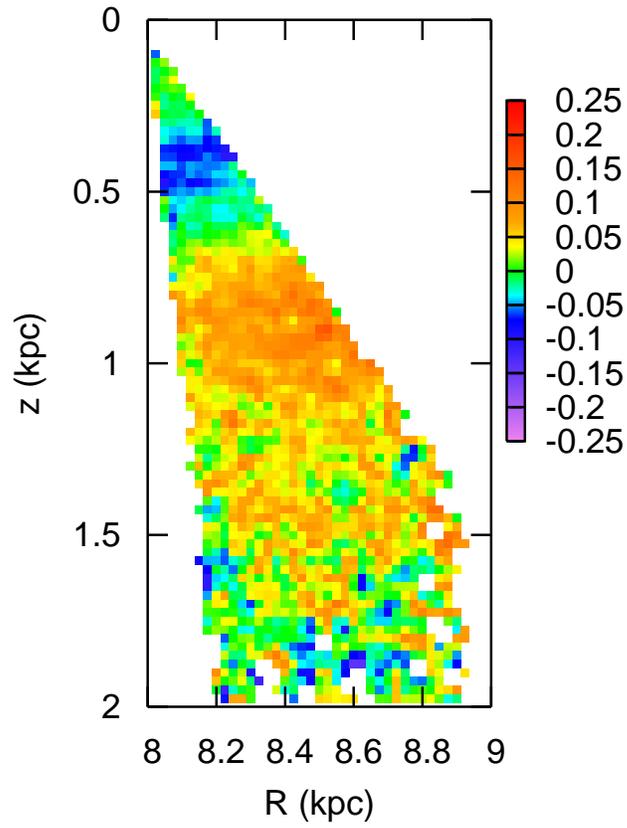}\label{fig:figure3}
\caption{North-South asymmetry $A_{2D}$ as a function of $z$ and the
  Galactocentric distance $R$.  We assume $8\,{\rm kpc}$ for the
  Sun-Galactocenter distance.}
\end{figure}

\begin{figure}
\epsscale{1.0}
\plotone{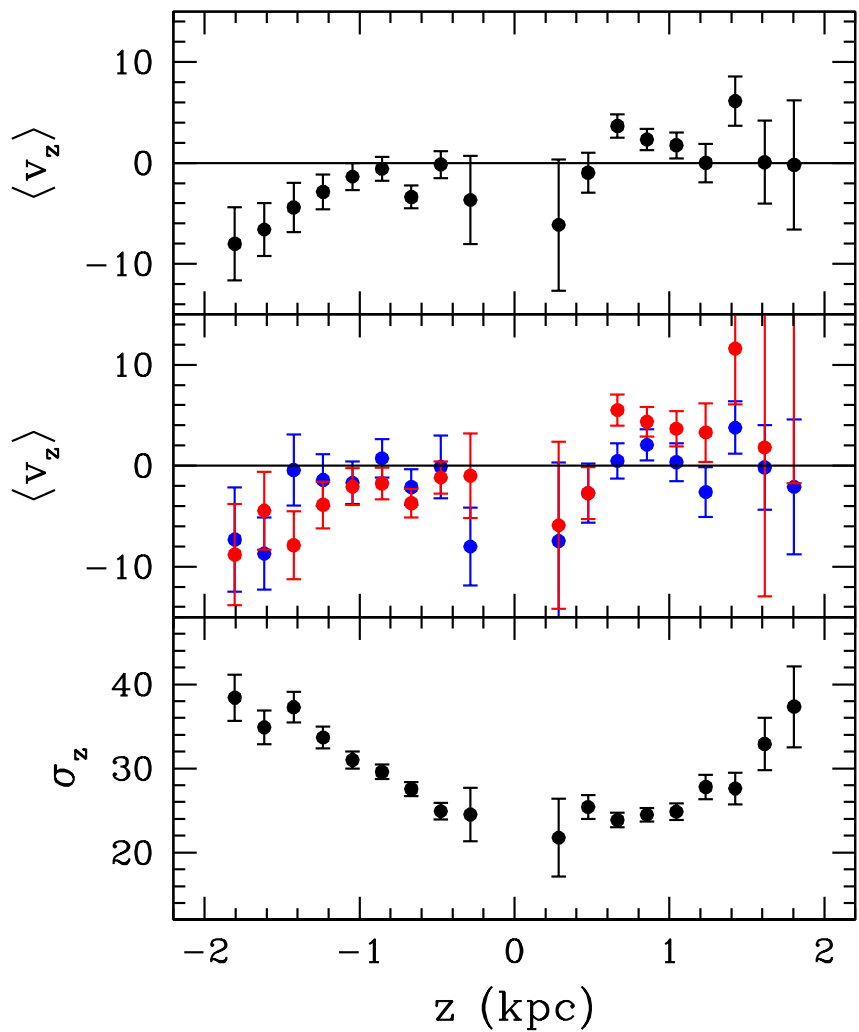}\label{fig:uvwns}
\caption{Bulk velocity $\langle v_z\rangle$ and velocity dispersion
 $\sigma_z$ as a function of $z$ in units of ${\rm km\,s}^{-1}$.  Top
 panel shows the bulk velocity as a function of $z$ for the entire
 spectroscopic sample.  Middle panel shows the bulk velocity profile
 for the ``red'' subsample ($g-r > 1$) and ``blue'' subsample
 ($g-r<1$).  The peculiar motion of the Sun ($v_{z,\odot} = 7.2\,{\rm
   km\,s^{-1}}$ \citep{dehnen98}) has been subtracted from $\langle
 v_z\rangle$.  Bottom panel shows the velocity dispersion.}
\end{figure}

\begin{figure}
\epsscale{1.0}
\plotone{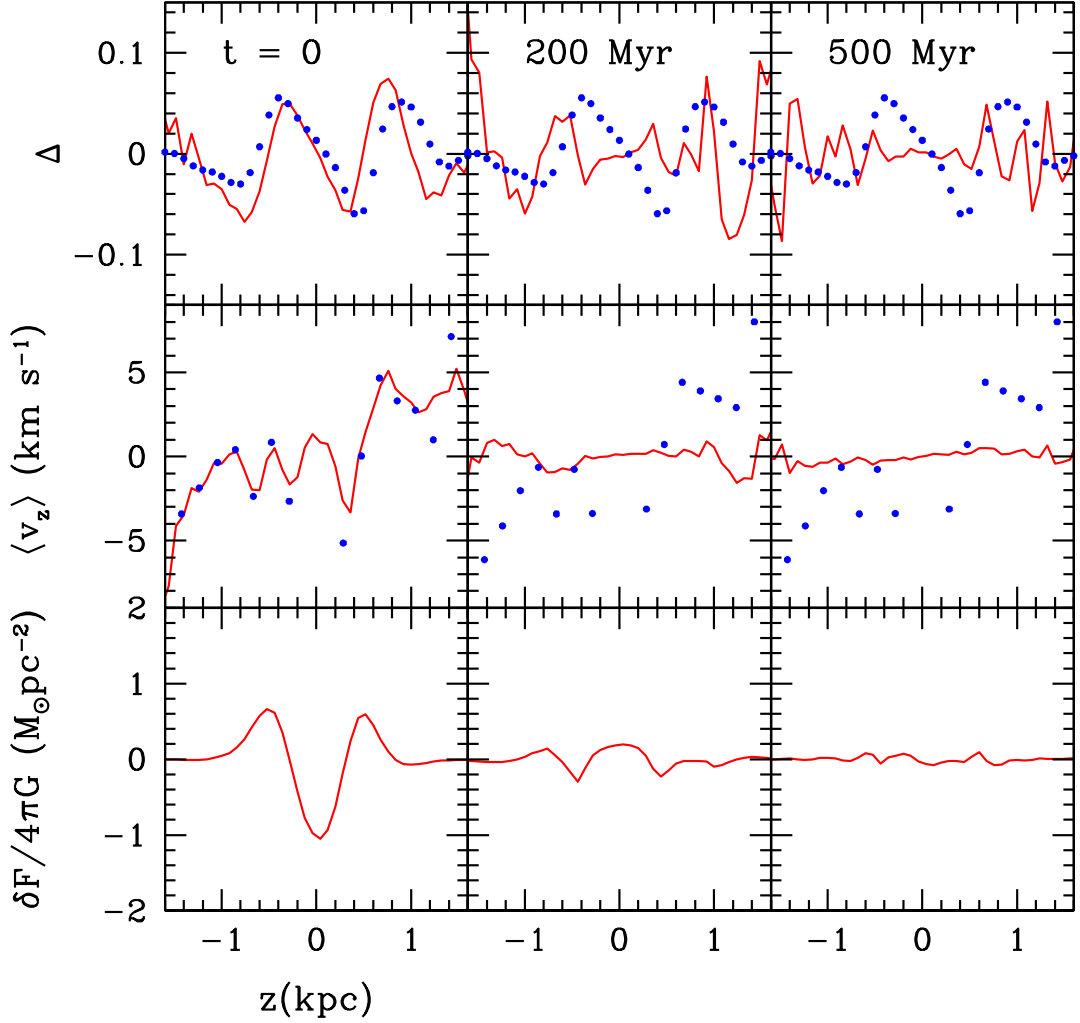}\label{fig:figure5}
\caption{Density, bulk velocity and vertical force as a function of
  $z$ for the initial conditions and for three time frames from the
  one-dimensional simulation described in the text.  Top panel shows
  $\Delta = \left (\rho(z) - \rho_0\right )/\rho_0$ where $\rho_0$ is the
  unperturbed density.  Middle panel shows the bulk velocity.  Bottom
  panel shows the difference between the vertical force for the perturbed and
unperturbed disk.  Blue points represent results from
  our analysis of the SDSS and SEGUE data.}
\end{figure}

\end{document}